\documentclass[fleqn,usenatbib]{mnras}

\usepackage{mathptmx}
\usepackage[T1]{fontenc}
\usepackage{ae,aecompl}

\usepackage{graphicx}	% Including figure files
\usepackage{amsmath}	% Advanced maths commands
\usepackage{amssymb}	% Extra maths symbols
\usepackage{graphicx,natbib}
\usepackage[colorinlistoftodos]{todonotes}
\usepackage{color}
\usepackage{rotating}
\usepackage{subfig}
\usepackage{txfonts}

\newcommand{\kms}{\,km\,s$^{-1}$}
\newcommand{\kpc}{\,kpc} % kiloparsec
\newcommand{\Msun}{\,M$_{\odot}$} % SolarMass
\newcommand{\degree}{$^{\circ}$}

\title[On the Origin of the Monoceros Ring]{On the Origin of the Monoceros Ring - I: \\
Kinematics, proper motions, and the nature of the progenitor}

\author[M. Guglielmo et al.]{Magda Guglielmo$^{1}$\thanks{e-mail: magda.guglielmo@sydney.edu.au},
Richard R. Lane$^{2,3}$,
Blair C. Conn$^{4}$,
Anna Y. Q. Ho$^{5}$, 
\newauthor
Rodrigo A. Ibata$^{6}$
and Geraint F. Lewis$^{1}$
\\
$^{1}$Sydney Institute for Astronomy, School of Physics, A28, The University of Sydney, NSW 2006, Australia\\
$^{2}$Millennium Institute of Astrophysics, Av. Vicu\~na Mackenna 4860, 782-0436 Macul, Santiago, Chile\\
$^{3}$Instituto de Astrof\'isica, Pontificia Universidad Cat\'olica de Chile, Av. Vicu\~na Mackenna 4860, Santiago, Chile\\
$^{4}$Research School of Astronomy and Astrophysics, Australian National University, Canberra, ACT 2611, Australia\\
$^{5}$Cahill Center for Astrophysics, California Institute of Technology, MC 249-17, 1200 E California Blvd, Pasadena, CA, 91125, USA\\
$^{6}$Observatoire astronomique de Strasbourg, Universit\'e de Strasbourg, CNRS, UMR 7550, 11 rue de l'Universit\'e, F-67000 Strasbourg, France
}

\date{Accepted XXX. Received YYY; in original form ZZZ}

\pubyear{2017}

\begin{document}
\label{firstpage}
\pagerange{\pageref{firstpage}--\pageref{lastpage}}
\maketitle

\begin{abstract}
The Monoceros Ring (MRi) structure is an apparent stellar overdensity that has been postulated to entirely encircle the Galactic plane and has been variously described as being due to line-of-sight effects of the Galactic warp and flare or of extragalactic origin (via accretion). Despite being intensely scrutinised in the literature for more than a decade, no studies to-date have been able to definitively uncover its origins. Here we use $N$-body simulations and a genetic algorithm to explore the parameter space for the initial position, orbital parameters and, for the first time, the final location of a satellite progenitor. We fit our models to the latest Pan-STARRS data to determine whether an accretion scenario is capable of producing an in-Plane ring-like structure matching the known parameters of the MRi. Our simulations produce streams that closely match the location, proper motion and kinematics of the MRi structure. However, we are not able to reproduce the mass estimates from earlier studies based on Pan-STARRS data. Furthermore, in contrast with earlier studies our best-fit models are those for progenitors on retrograde orbits. If the MRi was produced by satellite accretion, we find that its progenitor has an initial mass upper limit of $\sim10^{10}$\Msun~and the remnant is likely located behind the Galactic bulge, making it difficult to locate observationally. While our models produce realistic MRi-like structures we cannot definitively conclude that the MRi was produced by the accretion of a satellite galaxy.
\end{abstract}


%%%%%%%%%%%%%%%%%%%%%%%%%%%%%%%%%%%%%%%%%%%%%%%%%%

%%%%%%%%%%%%%%%%% BODY OF PAPER %%%%%%%%%%%%%%%%%%

\section{Introduction}
Surveys targeting the Milky Way (MW) disc during the past decade or so have uncovered many intriguing structures in its outer regions. With each of these discoveries questions naturally arise as to their origins and, as more of these structures are uncovered, how they relate to the origin and evolution of the Disc itself.

Arguably the most intriguing of these is a stellar overdensity apparently encircling the Disc dubbed the Monoceros Ring\footnote{Other names for this structure in the literature include: Galactic Anticenter Stellar Stream; Galactic Anticenter Stellar Structure; Monoceros Stream; Monoceros Overdensity} (MRi). This structure is particularly interesting because of its implications. If the MRi is of extragalactic origin then it is likely the remnants of an in-Plane accretion event, the only extant example of its type known in the Galaxy. In-plane accretion drives disc evolution by depositing stellar and gaseous material directly onto the disc, in a similar way to how the accretion of satellites on more polar orbits deposits material into the galactic halo \cite[as the Sagittarius dwarf galaxy,][does]{Ibata1994}.

First uncovered \cite[][]{Newberg2002} as a large stellar overdensity located $\sim18$\kpc~from the Galactic centre and with a scale height of $\sim2$\kpc\, and a scale length of $\sim10$\kpc, the MRi has subsequently been observed from 14-18 kpc from the Galactic centre over Galactic longitudes of 60\degree$<l<$280\degree and at distances from the Galactic plane of $|z|<5$\kpc~\cite[e.g.][]{Yanny2003,Ibata2003,Rocha2003,Conn2005a,Conn2005b,Conn2007,Conn2008,Conn2012,Sollima2011}.

Two explanations for the origin of the MRi have dominated the literature since it was uncovered \cite[see][for a detailed review of the history of studies of the MRi to-date]{Morganson2016}, namely a disrupting satellite and a Galactic origin scenario. The announcement of a possible dwarf galaxy in Canis Major \cite[][]{Martin2004} apparently boosted the possibility of it being formed by a disrupting satellite \cite[e.g.][]{Helmi2003,Martin2004,Penarrubia2005,Penarrubia2006,Martin2006,Conn2007,Sollima2011}. While in-plane accretion events could explain the origins of the MRi, it is still unclear whether the Canis Major region could host a dwarf galaxy remnant. \citet{2009AJ....137.4412M}, for example, reports that the Canis Major region has no overdensity of RR Lyrae stars as would be expected for a dwarf galaxy of this size. There is still debate over the origins of the Blue Plume stars in Canis Major ~\citep{2005ApJ...630L.153C} and no open clusters have been definitively associated with it. This does not exclude the possibility of an in-plane accretion event but does raise concerns about Canis Major being a dwarf galaxy and the progenitor the MRi.
%\todo[inline]{Geraint, Rodrigo, I'm putting the brakes on the CMa idea here and we have, in the past, put a lot of capital into it being a dwarf galaxy, are you happy with this position or would you prefer it being more ambiguous regarding its origins. I personally think its unlikely to be a dwarf galaxy, but I don't think standard milky way structure completely covers it either. -- Blair.}

As for Galactic origins, line-of-sight effects of the warp and flare of the Disc \cite[e.g.][]{Lopez2014,Kalberla2014}, and caustics in the dark matter profile of the Galaxy \cite[][]{Sikivie2003,Natarajan2007,Duffy2008} have also been investigated as reasons for the apparently ring-like stellar overdensity. More recently, apparent radial waves in the Disc observed in Sloan Digital Sky Survey \cite[SDSS: ][]{York2000} data were uncovered by \cite{Xu2015}.  These ripples may have been caused by, for example, interactions between the Galaxy and a Sagittarius-like dwarf satellite galaxy \cite[e.g.][]{Younger2008,Purcell2012,Gomez2013} and this has also been suggested as a solution to the MRi origin problem. Examination of these different scenarios reveal that the warp and flare models typically work for individual fields but generally have different parameters between fields and as such have no single solution for the entire structure. The other scenarios are either mostly qualitative or make predictions that are not seen in the data and consequently there is still no definitive explanation for the origin of the MRi.

It is also interesting to note that the rotation curve of the MW has a "kinematic dip" at $\sim9$\kpc~Galactocentric radius. The cause of this is unknown but a possible explanation is a massive stellar ring at this distance from the Galactic centre. The stellar ring, if it's existence can be confirmed, may be related to the Perseus spiral arm and is, therefore, likely not associated with the MRi \cite[][]{Sofue2009}. Furthermore, the Galactocentric distance of this purported stellar ring does not match well with the distance to the MRi and we only mention it here for completeness.

Recently \cite{Morganson2016} analysed Pan-STARRS-1 \cite[][]{Kaiser2010} data to estimate the three dimensional structure and stellar mass of the MRi. The region probed by the study where the MRi is most visible in the Pan-STARRS-1 dataset was $120$\degree$<l<240$\degree; $-30$\degree$<b<+40$\degree~and the total excess stellar mass was estimated as $4\times10^6$\Msun. However, despite the $3\pi$\degree~sky coverage and depth ($g\sim r\lesssim21.7$) of Pan-STARRS-1 the origins of the MRi remain unclear.

This work aims to reproduce the observed properties of the MRi. Using $N$-body simulations combined with a Genetic Algorithm, we 
%explore the possibility of \emph{an %extragalactic/accretion origin} for the MRi. 
examine the consequences of the accretion of a satellite to explore whether an overdensity of stars such as the MRi can form from such an accretion. The method enables a rigorous search of the parameter space with the selection of possible solutions by an automatic comparison between simulation results and data constraints, as described by \cite{Guglielmo2014} and, furthermore, allows for the current progenitor location to be estimated for the first time. In a subsequent paper we will address the rippled Disc hypothesis using similar methodology.

\section{Model}

\begin{figure*}
	\centering
	\includegraphics{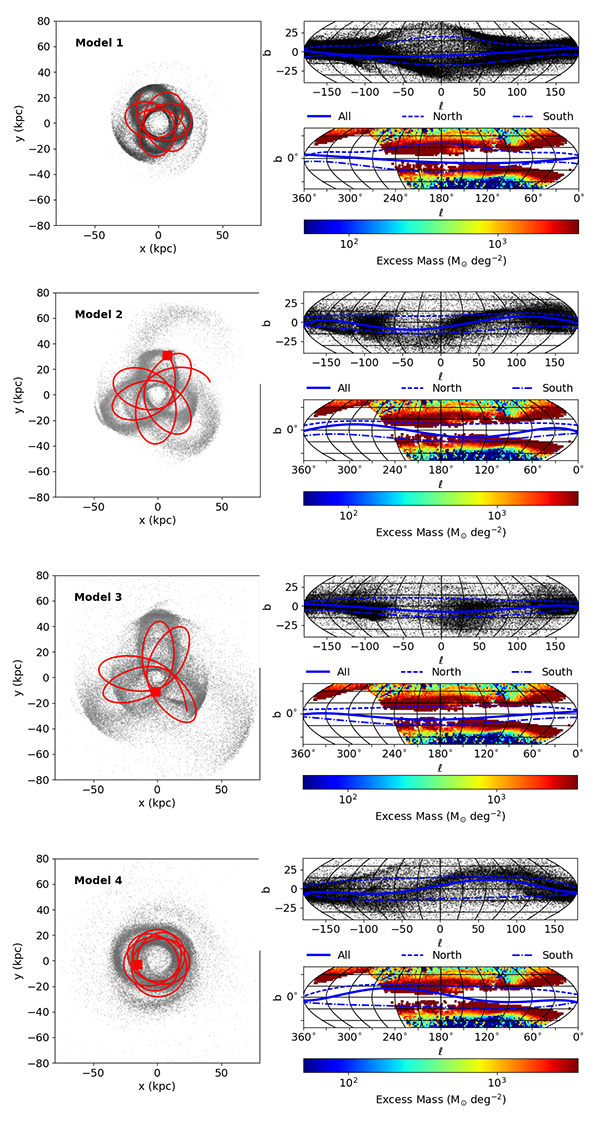}
	
	\caption{Results from a GA run on all three mass models (see \protect\ref{tab:SatMass}). \emph{ Left panels:}  Projection of particles in the $(x,y)$ plane after 3 Gyr. The solid red lines in all plots show the orbit of the progenitor,  whose present-day position is indicated by the solid red square. \emph{Right panels: Top:}  $(l,b)$ projection of the simulated stream in the region $-40^{\circ}<b<40^{\circ}$. The blue lines show the stream tracks for all particles (\emph{solid line}), for only those particles in the northern region (\emph{dashed line}) and only those in the southern region (\emph{dashed dot line}).  \emph{Bottom:} Comparison between the particle tracks with the data as presented by \protect\cite{Morganson2016}. For the four models, the values of the fitness function are $0.98$, $0.89$, $0.96$, $0.97$ for Models 1, 2, 3 and 4 respectively.} 
	\label{fig:Model_comparison}
\end{figure*}
\label{Models}

\subsection{$N$-body Simulations}

All of the simulations were carried out using the $N$-body code Gadget-2 \citep{Springel2005}. The original code has been modified to include the Milky Way gravitational influence on the progenitor. The equation of motion is, therefore, described by:
\begin{equation}
	\ddot{\textbf{r}}=\frac{\partial{\phi_{\rm{MW}}}(|\textbf{r}|)}{\partial{\textbf{r}}}+\frac{\textbf{F}_{\rm{df}}}{\rm{M_{sat}}}
\end{equation}
where $\phi_{\rm{MW}}$ is the total Galactic potential, $\rm{M_{sat}}$ is the satellite mass and $\textbf{F}_{\rm{df}}$ is the dynamical friction felt by the satellite when moving in the MW dark matter halo.
We describe the MW as a static, three-component potential, consisting of a   an exponential disk and a spherical bulge embedded into a dark matter spherical halo. The disk is modelled by a \cite{Miyamoto1975} potential, given by:
\begin{equation}
\phi_{D}(R,z)=-\frac{\rm{G\,M_{disk}}}{\left(R^2+\left( r_{\rm{disk}}+\sqrt{b^2+x^2} \right)^2\right)^{1/2} }.
\end{equation}

For the bulge component, we assumed a \cite{Hernquist1990} profile, described by:
\begin{equation}
\phi_{B}(r)=-\frac{\rm{G\,M_{bulge}}}{r_{\rm{bulge}}+r }.
\end{equation}
Finally, the dark mark matter halo follows a NFW profile \cite[][]{Navarro1997}, given by:
\begin{equation}
\phi_{H}(r)=-\frac{\rm{G\,M_{vir}}\,c}{\rm{R_{ vir}}\,g(c)} \ln{\left(\frac{r\rm{c}}{\rm{R_{ vir}}}+1\right)}
\end{equation}
where $c$ is the concentration parameter and $g(c)=\ln{(c+1)}-c/(c+1)$.

\begin{table}
	\centering
	\begin{tabular}{|l|r|}
		\hline
 	Parameter & Value\\
    \hline
    M$_{\rm{disk}}$ & $7.1\times{10^{10}}$\Msun  \\
    r$_{\rm{disk}}$ & $3.0$\kpc \\
    b$_{\rm{disk}}$ & $0.3$\kpc \\
    M$_{\rm{bulge}}$ & $0.7\times{10^{10}}$\Msun\\
    r$_{\rm{bulge}}$ & 2.1\kpc\\
    M$_{\rm{vir}}$ &$1.3\times{10^{12}}$\Msun  \\
    R$_{\rm{vir}}$ & 283 \kpc\\
    c  & 17\\
    \hline
    \end{tabular}
	\caption{Assumed parameters for the gravitational potential of the Milky Way}
	\label{tab:MWpar}
	\end{table}
 	
Table \ref{tab:MWpar} summarises the parameters used for the mass and radius of each component. They are constant through all our simulations. Our choice of parameters reproduces the current MW circular velocity  at the solar position ($R_{\odot}=8.29$ kpc), $V_{\rm{cir}}=239$\kms~\citep{McMillan2011} and are consistent with the recent estimation of the MW mass distribution \citep[e.g.][]{Kafle2014}.

The dynamical friction due to the dark matter halo is described by the Chandrasekhar equation \citep{Chandrasekhar1943}:
\begin{equation}
\textbf{F}_{\rm{df}}=-\frac{4{\pi}{\rm{G}^2}{\rm{M_{sat}}}{\ln{(\Lambda)}}\rho(\textbf{r}) }{v^2} \left[\rm{erf}(X)-\frac{2X}{\sqrt{\pi}}\exp{\left(-X^2\right)}\right]\frac{\textbf{v}}{v},
\end{equation}
where $\rho(\textbf{r}) $ is the density of the MW halo, $v$ is the orbital velocity of a satellite with mass M$_{\rm{sat}}$. The parameter $X=c/\sqrt{2\sigma^2}$ includes the velocity dispersion of the particles in the host halo. In this paper, we follow the analytical approximation for $\sigma$ for a NFW profile as described by \cite{Zentner2003}. The parameter $\ln(\Lambda)$ is the Coulomb logarithm and $\Lambda$ is defined here as the ratio between the radial position of the satellite and the impact parameter. This is chosen to be 2.0 kpc for Model 1, 2.5 kpc for Model 2 and 3.5 kpc for Model 3 and 4. 

\begin{table}
	\centering
	\begin{tabular}{|c|c|c|}
		\hline
		Model & M$_{\rm{DM}}$ ($10^{10}$\Msun) & M$_{\rm{*}}$ ($10^{10}$\Msun)  \\
		\hline
		Model 1 &0.03 & 0.003\\
		Model 2 &0.88 & 0.04\\
		Model 3 &1.13 & 0.05\\
		Model 4 &1.57 & 0.60\\
		\hline
	\end{tabular}
	\caption{Satellite Mass Models for Initial Conditions.}
	\label{tab:SatMass}
\end{table}

The influence of dynamical friction on the satellite orbits depends on the satellite mass and its velocity, two unknowns in our simulations. To overcome the lack of information on the mass, we have used three different mass models for the satellite as indicated in Table \ref{tab:SatMass}.
Early results, based on the Sloan Digital Sky Survey, suggested that the total stellar mass in the Monoceros Ring is between $2\times10^{8}$\Msun~and $\sim{10^{9}}$\Msun~\citep{Ibata2003}. Based on Pan-STARRS data, \cite{Morganson2016} estimate a total mass for the Monoceros ring in the range $[4,6]\times{10^{7}}$\Msun. %\todo[inline]{Magda, can you please tell me where this range came from? The best I can tell from the Morganson paper is their mass range is $4-6\times10^7$\Msun.}\todo[inline]{The abstract said that if you interpolate across the Galactic plane the mass is $8\times10^{6}$~\Msun while the observed excess mass is $4\times10^{6}$~\Msun. I assumed the interpolate mass as a good estimation for the minimum mass while $6\times10^{7}$~\Msun as maximum. Similarly, the conclusion. Anyway, I changed it now to your range.}
The stellar mass models used in our simulation span from $2\times{10^{8}}$ to $1\times{10^{9}}$\Msun. The dark matter halo mass changes according to the stellar mass, so that the baryonic fraction (stellar mass over dark matter mass) is approximately 0.07 in all models.

\subsection{Genetic Algorithm: Parameter Space and Orbit Selection}

Genetic Algorithms (GAs) are powerful population-based algorithms, able to solve optimisation problems by mimicking biological evolution. Starting from an initial \emph{population} of possible solutions (\emph{individuals}), the algorithm identifies those that better satisfy the model requirement. Each individual corresponds to a combination of free parameters, whose fitness depends upon the value of a merit function that describes the problem to convey. The best, or `fittest',  individuals are then used to generate a new set of solutions to form the next generation. At each step, the population evolves toward an optimal solution.

Free parameters in our GA are the past position and velocity of the progenitor, and they correspond to the only parameters used to identify the best orbital model. 
The $(x,y,z)$ positions are randomly selected between $-50$\kpc~and 50\kpc~from the Galactic Centre. The corresponding components of the total velocity are selected in such a way that 3\,Gyr ago the satellite was gravitationally bound to the Milky Way, or rather, that its total velocity was less than the MW escape velocity at that position. Within the $(x,y,z)$ range defined above the GA randomly selects  $(vx,vy,vz)$ between $-230$ and 230\kms.

Our populations consist of 50 possible solutions, each corresponding to particular initial conditions that are passed to the $N$-body code. After 3\,Gyr, the present day simulated snapshot is compared to observations. This comparison is done by firstly checking for the presence of stream-like structures. We define a `stream' as an over-dense structure of baryonic particles not gravitationally bound to the progenitor. The code selects all particles initially associated with the satellite stellar component and estimates the density of each based on the nearest particles. Stream candidates are those particles with a density greater than the total mean density.

If the simulation produces a stream, the code checks if it lies within the Monoceros Ring region. \cite{Morganson2016} characterised the MRi as  extending from $90^\circ<l<270^\circ$ with Northern ($3^\circ\lesssim b\lesssim45^\circ$) and Southern ($-45^\circ\lesssim b\lesssim-3^\circ$) components. Therefore, we require that the simulated streams cover a similar range of $l$ and present the characteristic Northern/Southern components. To do this the code considers the two components individually by creating a north and south distribution of particle positions in $(l,b)$. For each it calculates the interquartile range (IQR$_{\rm{sim}}$), as an indicator of the actual extent of the structure. IQR$_{\rm{sim}}$ is then compared with the equivalent value obtained from the data. These are taken from  the density map presented by \cite{Morganson2016} but focusing only on pixels with more than 100\Msun~excess mass. Each component is then compared with the data via:
%\begin{equation}
 \begin{align}
	F_{N,S}&=\frac{1.0}{1.0+\left(\frac{(\rm{IQR_{sim}})_{l_{N,S}}-(\rm{IQR_{data}})_{l_{N,S}}}{\sigma_{l}}\right)^2}\\
	 &\quad+\frac{1.0}{1.0+\left(\frac{(\rm{IQR_{sim}})_{b_{N,S}}-(\rm{IQR_{data}})_{b_{N,S}}}{\sigma_{b}}\right)^2}
\label{eq:fitness}
\end{align}
where $(\rm{IQR_{data}})_{l_{N}}=164.0^{\circ}$,   $(\rm{IQR_{data}})_{l_{S}}=141.0^{\circ}$, $\sigma_{l}=20^{\circ}$, $(\rm{IQR_{data}})_{b_{N}}=28.0^{\circ}$, $(\rm{IQR_{data}})_{b_{S}}=-20.0^{\circ}$, both with $\sigma_{b}=10^{\circ}$. The choice or $\sigma_l$ and $\sigma_{b}$ is arbitrary and used only to define a degree of similarity between the data and simulation. The final fitness function is defined as the product of each  component:
\begin{equation}
F=F_N*F_S
\label{eq:Fitness}
\end{equation}
and it is normalised to be 1 for a perfect match with observations. 

No further conditions are set for structures outside the defined region of the two components or on the orbit of the satellite. Optimal orbits are chosen based on the value for the fitness function (Equation \ref{eq:Fitness}).
%                                     Two column figure (place early!)
%______________________________________________ Gamma_1 (lg rho, lg e)
%
\section{Results}
\begin{table}
	\centering
	\begin{tabular}{|c|c|c|c|c|}
	\hline
 		Model & M$_{\rm{N}}$  &M$_{\rm{S}}$ & M$_{*f}(r<3$\kpc$\,)$\\
	 \hline
	 Model 1 & $(0.03\pm0.01)\times{10^8}$\Msun &$(0.04\pm0.01)\times{10^8}$\Msun & -\\
     Model 2 & $(0.8\pm0.2)\times{10^8}$\Msun &$(1.1\pm0.3)\times{10^8}$\Msun & $(0.3\pm0.3)\times{10^8}$\Msun \\
 Model 3 & $(1.5\pm0.5)\times{10^8}$\Msun &$(1.2\pm0.4)\times{10^8}$\Msun & $(0.3\pm0.5)\times{10^8}$\Msun \\
 Model 4 & $(1.7\pm0.8)\times{10^8}$\Msun  &$(1.2\pm0.4)\times{10^8}$\Msun & $(0.5\pm0.3)\times{10^8}$\Msun \\
  	\hline
\end{tabular}
\caption{This table lists the distribution of the stream mass in the Northern and South component; the progenitor final stellar mass distribution. The latter indicates the stellar mass within 3 kpc from the satellite centre (the lowest mass progenitor, Model 1, is totally destructed by the interaction with the Galaxy). The distributions are obtained by considering a sample of 1000 GA possible solutions with fitness function $>0.9$ (see text for more details). For all the four mass models, the results for the stream Northern and Southern components are not consistent with the mass given by \protect\citet{Morganson2016}. From observations, the Northern and Southern component have mass around $6-8\times{10^{6}}$\Msun and $4-5\times{10^{7}}$\Msun respectively,  while our simulations show a mass of a factor of ten time higher than the expected value.(See text for more details)}.
\label{tab:SimDetails}
\end{table}

For each mass model in Table \ref{tab:SatMass}, we study the evolution of an initial population of 50 individuals for a maximum of 50 generations. For each individual in the population, we study the interaction between the satellite and the MW for 3\,Gyr, starting from the past position randomly chosen by the GA as described in the previous section. The final snapshot corresponds to the present day position and it is used to evaluate the fitness function (Equation \ref{eq:Fitness}).

Figure \ref{fig:Model_comparison} shows an example of best solutions found by the GA for each mass model in Table \ref{tab:SatMass}. The specific mass models are indicated in the panels. The figure includes the final particle positions in the $(x,y)$ projection (left panels), with the progenitor orbits plotted as a red solid line. The left panels show the $(l,b)$ projection of the simulated streams in the Monoceros region ($-40^{\circ}<b<40^{\circ}$). In the top panels, the \emph{black points} show the particle distribution in $(l,b)$ as a result of the 3\,Gyr interaction, while the \emph{blue lines} show the average track of the stream for the Northern (\emph{dashed blue line}) and the Southern (\emph{dashed dot blue lines}) components and the for the entire stream  (\emph{solid blue lines}). In the bottom panels, the same line tracks are compared with the Pan-STARRS density \citep{Morganson2016}. As in \citeauthor{Morganson2016}, the map describes the total excess mass along the line of sight and it directly compares to their Figure 12. 

   As shown Figure \ref{fig:Model_comparison}, the $(l,b)$ projections of each GA run produce well-defined Northern and Southern structures similar to those observed in the Pan-STARRS data. These models have fitness values greater than $\sim0.89$, indicating that they reproduce the observed extent of the stream ($90^\circ<l<270^\circ$) in the north ($3^\circ\lesssim b\lesssim 45^\circ$) and south ($-45^\circ\lesssim b\lesssim -3^\circ$) fairly well.

The data presented by \cite{Morganson2016} allow for better constraints on the heliocentric distance of the MRi, showing that the southern component is closer ($d_{\odot}=6$~\kpc) than the northern one ($d_{\odot}=9$~\kpc). The models in Figure \ref{fig:Model_comparison} do not reproduce the distances of the two components, all producing streams at distances of 10\kpc~larger than expected. However, for Model 3 the simulations show a stream that is roughly 3\kpc~closer in the South than in the North ($d_{S}{_{\odot}}\sim15$\kpc~and $d_{N}{_{\odot}}\sim18$\kpc).

One interesting feature of the MRi overdensity is that the southern component appears to be more massive than its northern counterpart. \cite{Morganson2016} quote a mass for the Southern stream to be $4.8\times{10^7}$\Msun~and $8.6\times{10^6}$\Msun~for the Northern component. These are based on the assumption that the MRi is a uniform circle with a distance of 13\kpc~and 10\,\kpc~for the southern and northern stream respectively. The authors also note that the MRi centre (assuming a circular ring) might be off-centre by 4\kpc~with respect to the Galactic centre. This change in geometry led to a 30$\%$ difference in the mass estimation, (M$_{S}=3.3\times{10^7}$\Msun~and M$_{N}=6.0\times{10^6}$\Msun). However, despite the change of the system geometry, the general conclusion is that the stream is more massive in the south.

To analyse the GA results we build a sample of 1000 possible solutions all with a fitness function greater than 0.9. For each solution we estimate the mass of each component as well as the stellar mass of the progenitor within a radius of 3\,\kpc~from its centre. Among the four models, only the progenitor with mass $~3\times{10^8}$\Msun, (Model 1), does not survive the 3 Gyr interaction with the Milky Way. In Table \ref{tab:SimDetails}, we summarise the results of our statistical analysis indicating the mass value for the Northern and Southern components (column one and two, respectively) and the stellar mass of the progenitor remnant when possible. 

In general the simulations tend to overestimate the mass of the stream components resulting in a total mass of at least ten times higher than the value suggested by observations. The overestimation is more pronounced for Model 2 and Model 3 than Model 1, for which the progenitor has a lower initial stellar mass (see Table \ref{tab:SatMass}). Between the three mass models, Model 1 is the only one that can reproduce a Southern stream more massive than its northern counterpart.

 \subsection{Progenitor location}
 
Independent of the progenitor mass and its orbit, our simulations suggest that the progenitor survived the interaction with the Galaxy, although with a consistent mass loss.

As discussed in the previous section, the last column
in Table \ref{tab:SimDetails}  lists the final stellar mass within 3 kpc of the progenitor centre of mass.  The simulations suggest that the remnant galaxy has a stellar mass of the order of $10^{7}$~\Msun, although these estimations have significant uncertainties.

To identify the current progenitor location, we select a sample of 1000 possible orbits in which the satellite galaxy survived after a 3\,Gyr interaction with the Milky Way. These orbits are the results of repeated GA runs using all the three mass models. In each GA run, we identified and selected a variety of orbital models with fitness value greater than 0.9 to guarantee the formation of the MRi structure. The selected orbits do not necessarily correspond to the last generation, but we extend our selection to all generations. In other words, our samples contain all orbital models able to reproduce the MRi to a reasonable precision. 

\begin{figure}
\centering
%
%\subfloat
%	{
%		\includegraphics[scale=0.4]{M1_all.png}
%}\\
%\subfloat{
%		\includegraphics[scale=0.4]{M2_all.png}
%	}\\
%\subfloat{
%		\includegraphics[scale=0.4]{M3_all.png}
%	}
		\includegraphics{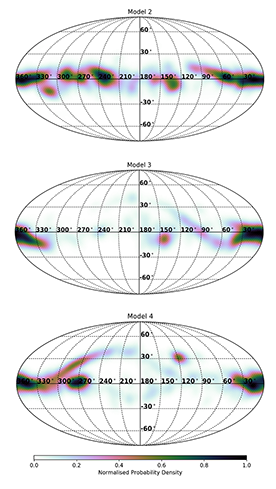}

\caption{Mollweide Projection Map of all the possible progenitor locations in Galactic coordinates ({\it l,b}), for Models 2, 3 and 4. A satellite with mass $\sim10^{8}$\Msun~(Model 1) does not survive the interaction with the Galaxy, hence there is no progenitor location at the end of the simulation. We consider 1000 possible orbital scenarios in each of the GA runs where the progenitor remnant is extant at the end of the simulation (after 3\,Gyr). The colour map is chosen so that darker regions represent locations of higher probability.}

\label{fig:m2prog}%
\end{figure}

Figure \ref{fig:m2prog} shows a density map of all the possible progenitor locations resulting from our analysis. The colour scale is such that the green/dark regions correspond to the location with higher occurrences. For Model 2 (\emph{top panel}), the most likely region for progenitor locations spans over a large longitude range. However, it is possible to identify three main regions around $(l,b)=(14,-1)$\degree~(with $99\%$ probability), $(l,b)=(352,-2)$\degree~($97\%$ probability) and $(l,b)=(232,2)$\degree~($90\%$ probability). For Model 3 (\emph{middle panel}) the most likely region for the progenitor location appears to be around $(l,b)=(11,-2)$\degree~($99\%$) and $(l,b)=(354,-1)$\degree~($97\%$). Similarly for the more massive model, Model 4 (\emph{bottom panel}), the progenitor remnant  seems to be confined in three region with comparable probability:  $(l,b)=(17,-2)$\degree~($99\%$), $(l,b)=(249,-1)$\degree~($98\%$) and $(l,b)=(271,2)$\degree~($93\%$). 
By comparing the three maps in Figure \ref{fig:m2prog}, it is interesting that all the three models indicate the region around $(l,b)\sim(15,0)$\degree~as the best candidate for the progenitor location. 

\begin{figure}
	\centering
	\includegraphics[scale=0.6]{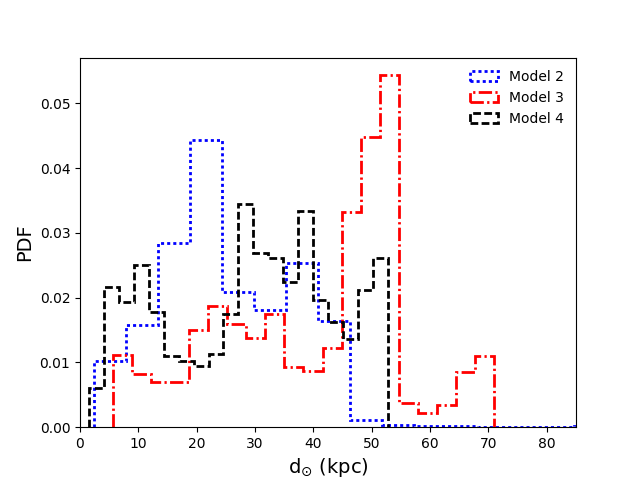}
\caption{Distribution of the progenitor's heliocentric distance for  Model 2 (\emph{red dashed line}), Model 3 (\emph{black dashed line}) and Model 4 (\emph{grey solid line})}
\label{fig:Distance}
\end{figure}

Figure \ref{fig:Distance} shows the distribution of the heliocentric distances for Model 2 (\emph{blue solid line}) Model 3 (\emph{red dashed line}) and Model 4 (\emph{black dotted line}). As for the location in Galactic coordinates, it is difficult to constrain the distance of the progenitor due to the large scatter of its final location. However, the three distributions have similar mean values of $27\pm12$\kpc, $40\pm12$\kpc~and  $31\pm13$\kpc~for Models 2, 3 and 4, respectively.


\subsection{Stream Kinematics}

\begin{figure*}
	\centering
	\includegraphics[scale=0.8]{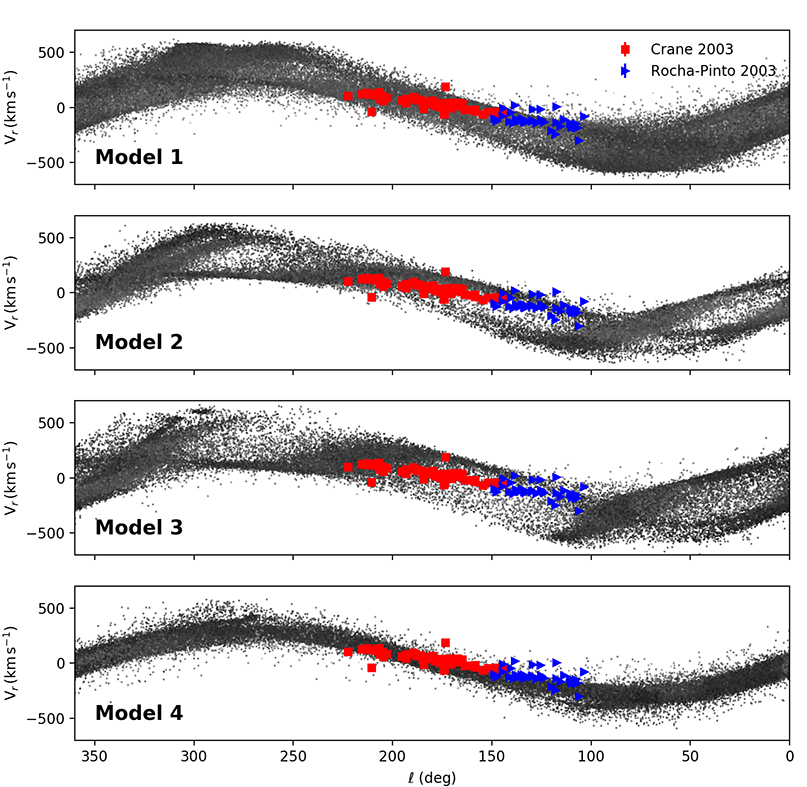}
	\caption{Heliocentric Radial Velocity versus longitude for the four models, as indicated. The radial velocities of the simulated streams are compared with the data from \protect\cite{Crane2003} (\emph{red squares}) and  \protect \cite{Rocha2003} ((\emph{blue triangles})). For better comparison with observations, Sun distance and Milky Way circular velocity are 8.5 \kpc and 220 \kms.}
	\label{fig:vrad}
\end{figure*}

\begin{figure*}
	\centering
%    \subfloat
%	{
%		\includegraphics{mbvsb.png}
%	}
%    \subfloat
%	{
%		\includegraphics{mlvsb.png}
%	}
	\includegraphics{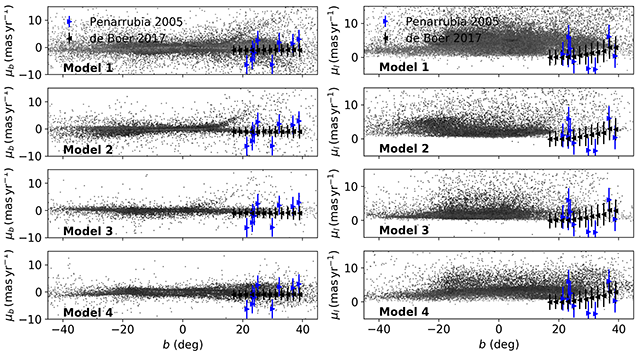}
	\caption{Proper motions in the latitudinal (\emph{left panels}) and in the longitudinal (\emph{right panels}) components as a function of latitude for the four mass models in Figure \ref{fig:Model_comparison}. In each panel, the blue solid triangles are the corresponding proper motion components of ten stars in the MRi identified by \protect{\citep{Penarrubia2005}}, with an error of 3.5 mas\,yr$^{-1}$. The red solid triangles show the proper motion values from \emph{Gaia} DR1 as described in  \protect{\citep{deBoer2017}}, with error bars equal to the dispersion listed in their Table 1. For the simulated streams, the black points indicate only those particles in the region $90$\degree$<l<240$\degree, for better comparison with osservations.}
	\label{fig:pm}
\end{figure*}

In this section we analyse the kinematic distribution for the three orbital scenarios described in Figure \ref{fig:Model_comparison}. Figures \ref{fig:vrad} and \ref{fig:pm} together describe the kinematic properties of the simulated stream for each mass model used in this work.  

Figure \ref{fig:vrad} describes the heliocentric radial velocity of the stream particles. We compare our simulations (\emph{grey points}) with the observational data by \cite{Crane2003} (\emph{red squares}) and \citep{Rocha2003} (\emph{blue triangles}), both based on the Two Micron All Sky Survey (2MASS) database. It is clear from this figure that the models well match the gradient of the radial velocity curve seen in the data, regardless the initial mass used for the satellite or its orbits. This is not surprising as different authors have shown that models with different orbital parameters, such as eccentricities or rotational sense of motion, can reproduce the observed radial velocity \citep{Martin2004,Penarrubia2005}.

A more complete picture of the kinematic properties of the stream is provided by the proper motions of the stars in the MRi. In Figure \ref{fig:pm}, we show the latitudinal (\emph{left panels}) and longitudinal (\emph{right panels}) the proper motion components of all particles in the system as function of their latitude. In each panel, the  proper motions of all particles in the MRi region ($90^{\circ}<\ell<240^{\circ}$) are compared with those of ten confirmed stellar members (\emph{blue squares}) presented in \cite{Penarrubia2005} (see their Table 2). The \emph{solid red triangles} show the stellar proper motion of the MRi obtained by combining SDSS astrometry and positions from \emph{Gaia} DR1 \citep{deBoer2017}. As for Figure \ref{fig:vrad}, all models well reproduce within the errors the stellar proper motions.

The comparison between the observed proper motions and simulations is crucial in understanding the direction, retrograde or prograde, of the progenitor orbit.  As discussed in Section \ref{Models}, our selection of the best models is based only on the ability of each orbit to reproduce the spatial properties of the Monoceros Ring. Therefore, there is no preference or distinction between the orbital sense of motion. Following the definition in  \cite{Penarrubia2005}, we defined a prograde orbit if the  orbital inclination $i$ is less than 90$^\circ$ and retrograde if $i$ is in the range  90$^\circ$ and 180$^\circ$. The orbital inclination $i$ is calculated as:
\begin{equation}
\cos({i})=-\frac{L_z}{L}
\end{equation}
where $L=\sqrt{L{^2}_{x}+L{^2}_{y}+L{^2}_{z}}$ is the total  angular momentum per unit of mass. In all the models in Figure \ref{fig:Model_comparison}, the progenitor lies on a retrograde orbit, with an  inclination of $\approx160^\circ$ for all three models.  

\begin{figure*}
 	\centering
%    \subfloat
%	{\includegraphics{M1_ret.png}}
%   \subfloat
%	{\includegraphics{M1_pro.png}}
	\includegraphics{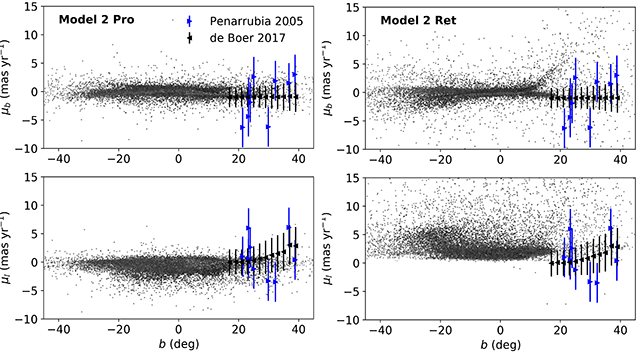}
 
 \caption{Comparison between retrograde (\emph{left panels}) and prograde (\emph{right panels}) orbit for Model 2, corresponding to a progenitor mass of $0.9\times10^{10}$ \Msun. The latitudinal (\emph{top}) and longitudinal (\emph{bottom}) proper motion components are compared with those by \protect{\citet{Penarrubia2005}} {\emph{(blue filled triangles)}} and data by \protect{\citet{deBoer2017}} {\emph{(black filled triangles)}}. Both orbital models reproduce the $\mu_b$ component of the proper motions well, but both fail in reproducing the $\mu_l$ data. As for Figure \ref{fig:pm}, the black points are only those particles in the region $90$\degree$<l<240$\degree, for better comparison with observations.}
%\begin{figure*}
%	\centering
%	\includegraphics[scale=0.5]{M1_kin.png}
%	\caption{Comparison between retrograde (\emph{left panels}) and prograde (\emph{right panels}) orbit Model 1 corresponding to a progenitor mass of $0.9\times10^{10}$ \Msun. \emph{From top:} Radial velocity as function of longitude compared with data from \citep{Crane2003} (\emph{red squares}) and \protect \citep{Rocha2003} (\emph{blue triangles}); latitudinal (\emph{middle}) and longitudinal (\emph{bottom}) proper motions components compared with \protect{\citep{Penarrubia2005}}}.
\label{fig:M1_kin}
\end{figure*}

In  Figure \ref{fig:M1_kin},  we compare  kinematic properties of the Model 2 retrograde orbit in Figure \ref{fig:pm} with one of its prograde equivalent (similar fitness function). Independently from their direction of motion, both models reproduce the latitudinal component of the proper motions. However, the two models produce two different ranges for the $\mu_l$ component of  proper motion. In the  stream region ($90{^{\circ}}\le l\le{240^{\circ}}$), for the prograde orbit (\emph{right panels}), $\mu_l$ spans from a minimum value of -2 mas$\,$yr$^{-1}$ to a maximum of 0 mas$\,$yr$^{-1}$ and for the retrograde orbit  (\emph{left panels}), $\mu_l$ ranges between 2\,mas$\,$yr$^{-1}$ and 5\,mas$\,$yr$^{-1}${\footnote{For the retrograde orbits, the range of the longitudinal component refers to the most dense region of the stream and ignores the more disperse particles}}, while from observations in \cite{Penarrubia2005} $\mu_l\in (-7,10)$\,mas$\,$yr$^{-1}$, including a 3.5\,mas$\,$yr$^{-1}$ uncertainty. Although neither model reproduces the observed range of proper motions in full, the retrograde orbital scenario seems better encompasses the observed data points.

\subsection{Comparison with data and previous models}

As discussed in previous sections, \cite{Penarrubia2005} presented the first model for the MRi, under the assumption that it formed by the disruption of an accreted satellite. The approach by \citeauthor{Penarrubia2005} is not dissimilar to the one presented here, as both methods used the available observational data to constrain the orbital properties of the MRi. However, the combination of a GA and $N$-body simulations allows us to select the optimal models through a direct comparison between simulations and observations while exploring a large parameter space. This difference implies that there are no \emph{a priori} conditions on the orbital parameters or the progenitor orbital directions.  

One of the primary results by \cite{Penarrubia2005} is that prograde orbits are favoured over retrograde ones, as the latter fail in reproducing (in magnitude and sign) the proper motion of 10 confirmed stars members of MRi. In addition, their retrograde orbits produce a much larger azimuthal angular velocity ($\mu_l\in{(5,20)}$ mas\,yr$^{-1}$) than the observations. These conclusions seem to be supported by more recent observations. Proper motion data obtained using Gaia DR1 seem to favour a prograde orbit \cite{deBoer2017}. On the other hand, \cite{Sheffield2014} conclude that a model with a progenitor satellite on a retrograde orbit can better reproduce the two distinct populations at the same orbital phase observed in 2MASS data.

When performing a similar analysis to the one presented by \cite{Penarrubia2005} (see Figures \ref{fig:pm} and \ref{fig:M1_kin}), our retrograde and prograde orbits can both reproduce the observed proper motion range equally. From our analysis, we conclude that it is difficult to discriminate the sign of the progenitor orbital motion from our simulations. However, it is necessary to clarify that the GA does not use proper motion constraints during the orbit selections. Introducing a further condition to the fitness function (Equation \ref{eq:fitness}) to match the proper motions of stars in the stream will allow  for a more  restrictive selection of the orbital models, and hence, a better selection of the orbital sense of motion.

Due to the lack of complete and deep data sets near the Galactic plane, the estimate of the total mass of this structure poses a challenge. Using data taken with Isaac Newton Telescope Wide Field Camera, \cite{Ibata2003}  predicted a total stellar mass in the structure to be between $\sim2\times{10^8}$\Msun~and ${10^9}$\Msun. This estimate was based on the assumption that the MRi is smooth and axisymmetric. By interpolating across the unobserved regions in the Pan-STARRS data, \cite{Morganson2016} provide constraints on the mass of the MRi. \citeauthor{Morganson2016} showed that the two components of the stream appear to have different masses, with the Southern overdensity ($M_S=4-5\times{10^7}$~\Msun~) being more massive than its northern counterpart ($M_N=6-9\times{10^6}$\Msun). In addition, they indicate a total mass of ${10^7}$~\Msun~as a lower limit.
 
As inferred from Table \ref{tab:SimDetails}, our models overestimate the total mass of the stream by a factor 10, and indicate equal masses for the Southern and Northern components. Only the lower mass model (Model 1) can reproduce the mass difference between the two components, albeit both of our Model 1 components have a mass smaller than the corresponding observed one. 

Even considering the possibility that \cite{Morganson2016}  underestimate the total mass and/or the mass of the two components, the results obtained by our Model 3 and Model 4 seem to be too high compared with observations, although the stream masses  are within the limit predicted by \cite{Ibata2003} (with models giving a total stream mass of $2-3\times{10^8}$ \Msun).  On the other hand, the two models with low mass, Model 1 and Model 2,  can better reproduce the mass of the two components. These results can fix an upper limit to the mass of the progenitor to be no higher than $\approx{1\times10^{10}}$\Msun.

%______________________________________________________________

\section{Summary and Conclusions}

We have used $N$-body simulations in combination with a Genetic Algorithm to explore the parameter space for the initial position, orbit and final location of the progenitor to efficiently probe the possibility of an accretion origin for the Monoceros Ring. The power of this work is in the GA which allows for a best fit to the latest data and for the $first~time$ allows for the current location of the progenitor to be estimated, something so far lacking in the literature.

The initial location and velocity of the modelled MRi progenitor are assigned as free parameters, with the location randomly selected by the GA between $-50$\kpc~to $50$\kpc~from the Galactic centre and the velocity between $-230$ and $230$\kms. The progenitor is assumed to be bound to the Galaxy after 3\,Gyr of evolution using the $N$-body code Gadget-2 (assumed Galactic mass parameters are shown in Table \ref{tab:MWpar}). The fitness (Equation \ref{eq:Fitness}) of the final snapshot, based on comparison with PanSTARRS-1 data, is calculated.

We present here three mass models (Table \ref{tab:SatMass}). For Model 1 we find an MRi mass of $\sim10^{8}$\Msun~and our Models 2 \& 3 produce an MRi mass of $\sim3\times10^8$\Msun. \cite{Morganson2016} estimate the mass to be $\sim4-6\times10^7$\Msun. This means our models contain {\emph at least} $\sim$10 times the mass estimated by \cite{Morganson2016} from the Pan-STARRS dataset. This presents a problem for the accretion scenario, however, we have demonstrated, in principle, that it is possible to reproduce a Northern stream that is less massive than the Southern stream as found by \cite{Morganson2016}. Even considering the possibility that \cite{Morganson2016} underestimated the total mass of the MRi and/or its two components, our Models 2 \& 3 seem to be too high. However, our modelled stream masses are within the limits predicted by \cite{Ibata2003}. Because our Model 1 (the model with the lowest progenitor mass) is our only model that produced a Northern stream roughly 10 times less massive than the Southern stream we can use this to fix the upper limit of the progenitor to $\sim10^{10}$\Msun~ (under the assumption that the MRi was produced via satellite accretion).% The final heliocentric distance for the progenitor is $\sim35$\kpc~for Models 1 and 2 and $\sim9$\kpc~for Model 3. This exposes the sensitivity of the final stream morphology to the assumed initial mass of the progenitor.

We estimate the final location of the progenitor for each of the 1000 realisations of each mass model with the results presented for our three mass models in Figure \ref{fig:m2prog}. For all three models we find a large range of final progenitor positions, making it very difficult to estimate a likely final position. However, it is possible to identify regions of the highest likelihood. For Model 1 we identify three regions around $(l,b)=(14,1)$\degree~(99\% probability), $(l,b)=(352,-2)$\degree~(97\% probability) and $(l,b)=(232,2)$\degree~(90\% probability). For Model 2 we can identify two regions at $(l,b)=(11,-2)$\degree~(99\%) and $(l,b)=(354,-1)$\degree~(97\%) and for Model 3 $(l,b)=(17,-2)$\degree~(99\%), $(l,b)=(249,-1)$\degree~(98\%) and $(l,b)=(271,2)$\degree~(93\%). Interestingly, all three models have a high likelihood of finding the progenitor around $(l,b)\sim(15,0)$\degree. Similarly we find the modelled Galactocentric distances clustered around $27\pm12$\kpc, $40\pm12$\kpc and $31\pm13$\kpc for Models 1, 2 and 3, respectively. If we boldly assume that the MRi really does have an accretion origin, despite its high mass ($\sim10^8$\Msun) the location of progenitor at $(l,b)\sim(15.0)$ would (somewhat conveniently) make it very difficult to locate observationally due to its location behind the Bulge.

While our Model 1 gives the closest mass estimate to that by \cite{Morganson2016} and to the radial velocities by \cite{Crane2003,Rocha2003}~(Figure \ref{fig:vrad}), it appears that, at least qualitatively, our Model 3 best recovers the proper motions (Figures \ref{fig:pm} \& \ref {fig:M1_kin}).%clear that the spread in the measured heliocentric radial velocities is best reproduced by our Model 1.
~However, it should be noted that we did not include radial velocity in our GA fitting parameters so we have not done any fitting to radial velocity. Even so, the gradient of the radial velocities with Galactic longitude is well described by all three of our mass models. This shows at the very least that an accretion scenario can reproduce the observed heliocentric radial velocity profile of the MRi \cite[as also implied by the ][studies]{Martin2004,Penarrubia2005}.

%Comparison with the proper motions of the 10 confirmed MRi "member" stars \cite[Table 2 by][]{Penarrubia2005} shows that again, qualitatively, all three models reproduce these kinematics (Figure \ref{fig:pm}) equally well. 
As with the radial velocities the proper motions were not included in the fitting procedure in our models. This is further evidence that the kinematics of the MRi can be reproduced with an accretion scenario. It should be noted that in contradiction with \cite{Penarrubia2005} the best fit models to the Pan-STARRS-1 data from all three of our mass models are all from progenitors on retrograde orbits. Those authors favoured prograde orbits in their models. Note that all of our models reproduce both the radial velocity gradient of the stream and the latitudinal component of the proper motions. However, our models produce two different ranges for the $\mu_l$ component of  proper motion. In the region ($90{^{\circ}}\le l\le{240^{\circ}}$), for the prograde orbit $\mu_l$ ranges from a minimum of -2\,mas$\,$yr$^{-1}$ to a maximum of 0\,mas$\,$yr$^{-1}$ and for the retrograde orbit $\mu_l$ ranges between 2\,mas$\,$yr$^{-1}$ and 5\,mas$\,$yr$^{-1}$, while from observations $\mu_l\in (-7,10)$\,mas$\,$yr$^{-1}$ (including a 3.5\,mas$\,$yr$^{-1}$ uncertainty). Contrary to early models \citep{Penarrubia2005, Sheffield2014}, we can not easily discern the favourite orbital model of the progenitor from our simulations. When compared with the observed proper motions, retrograde and prograde orbits reproduce the observations equally well (see Figure \ref{fig:M1_kin}). However, the current version of the GA does not include any comparison between model and observed proper motions. Including such comparison in the fitness function will help in identifying the preferred orbital orientation of the progenitor. The future data release of \emph{Gaia} will provide crucial information on the proper motions that will offer tighter constraints on the fit that will lead to better models.

%While neither model reproduces the observed range of proper motions in full, the retrograde orbital scenario better encompasses the observed data points (see Figure \ref{fig:M1_kin}).

The combination of a GA and $N$-body simulations has allowed us to select the optimal models through a direct comparison between simulations and observations by exploring a large parameter space. Our models imply that there are no \emph{a priori} conditions on the orbital parameters or the progenitor orbital directions, however, our model analyses lead to different conclusions that those presented by \cite{Penarrubia2005} regarding the orbital direction of the progenitor.

Our models demonstrate that an accretion origin for the MRi is not excluded, but we cannot state definitively that this is the favoured scenario. Here we have explored an extragalactic/accretion origin for the MRi and, therefore, cannot discount the possibility that the observed stellar overdensity dubbed the MRi has a different origin. In a subsequent paper we will address the possibility of the rippled Disc hypothesis using a similar methodology.

\section*{Acknowledgements}
RRL acknowledges support by the Chilean Ministry of Economy, Development, and Tourism's Millennium Science Initiative through grant IC120009, awarded to The Millennium Institute of Astrophysics (MAS). RRL also acknowledges support from the STFC/Newton Fund ST/M007995/1 and the CONICYT/Newton Fund DPI20140114. BCC acknowledges the support of the Australian Research Council through Discovery project DP150100862. A.Y.Q.H. was supported by a National Science Foundation Graduate Research Fellowship under Grant No. DGE-1144469.
The authors acknowledge the University of Sydney HPC service at The University of Sydney for providing HPC resources that have contributed to the research results reported within this paper.

%%%%%%%%%%%%%%%%%%%%%%%%%%%%%%%%%%%%%%%%%%%%%%%%%%



%\bibliographystyle{mnras}


%%%%%%%%%%%%%%%%%%%%%%%%%%%%%%%%%%%%%%%%%%%%%%%%%%

%%%%%%%%%%%%%%%%% APPENDICES %%%%%%%%%%%%%%%%%%%%%

%\appendix



%%%%%%%%%%%%%%%%%%%%%%%%%%%%%%%%%%%%%%%%%%%%%%%%%%

% Don't change these lines
\bsp	% typesetting comment
\label{lastpage}
\end{document}